\documentclass[11pt,twoside,a4paper,cr,final]{cms-tdr}
\def\svnVersion{67509}

\begin{document}

\hyphenation{had-ron-i-za-tion}
\hyphenation{cal-or-i-me-ter}
\hyphenation{de-vices}

\RCS$Revision: 62698 $
\RCS$HeadURL: svn+ssh://svn.cern.ch/reps/tdr2/papers/XXX-08-000/trunk/XXX-08-000.tex $
\RCS$Id: XXX-08-000.tex 62698 2011-06-21 00:28:58Z alverson $
\cmsNoteHeader{CR-2011/091} 
\title{Isolated photon production in $\sqrt{s_{NN}}$ = 2.76 TeV PbPb collisions as a function of transverse energy and reaction centrality}

\author[mit]{Yongsun Kim for the CMS collaboration}

\date{\today}

\abstract{
In studies of the dense medium produced in ultra-relativistic heavy ion collisions, photons are important hard probes, since they are not expected to be modified by the medium. The measurement of isolated prompt photon production in PbPb collisions provides a test of perturbative quantum chromodynamics (pQCD) and the information to constrain the nuclear parton distribution functions. CMS has shown  photon purity measurement capabilities in pp collisions using the shower shape templates.
In PbPb co llisions at CMS, this technique was applied for the first time in heavy ion collisions.  We report the first measurement of the transverse momentum spectra of isolated photons with pT from 20 GeV/c to 80 GeV/c in PbPb collisions at $\sqrt{s_{NN}}$ =2.76 TeV.  The centrality dependence of the nuclear modification factor is also reported by comparing the result to the photon spectrum of pp reference which is computed from NLO calculations. 

\vspace{8mm}

\begin{center}
Presented at \textit{QM2011}: \textit{Quark Matter 2011}
\end{center}
}

\newcommand {\XT}          {\ensuremath{x_T}}
\newcommand{\RAA}        {\ensuremath{R_{AA}}}
\newcommand{\TAA}        {\ensuremath{T_{AA}}}
\newcommand{\Npart}      {\ensuremath{N_{part}}}
\newcommand{\Ncoll}       {\ensuremath{N_{coll}}}

\newcommand{\microbinv} {\mbox{\ensuremath{\,\mu\text{b}^\text{$-$1}}}\xspace}

\hypersetup{%
}

\maketitle

\section{Introduction}

High transverse energy ($E_T$) prompt photons in nucleus-nucleus collisions
are produced directly from the hard scattering of two partons.
Once produced, photons traverse the produced hot and dense medium without interacting strongly, they provide a direct test of perturbative QCD (pQCD) and the nuclear parton densities~\cite{Arleo:2011gc}.
In this analysis, we report the isolated photon production in PbPb collisions at $\sqrt{s_{NN}} = 2.76 $ TeV as a function of the event centrality with the CMS detector~\cite{JINST}.

\section{Analysis methods}

The events for this analysis are selected by requiring two triggers to be fired, a Level-1 (L1) electromagnetic cluster with $E_T> 5$ GeV and a High Level Trigger (HLT) photon
with $E_T > 15$ GeV.
The efficiency is $> 98\%$ for photon candidates with corrected transverse energy $E_T > 20$ GeV.
In addition to the photon data sample, a minimum-bias (MinBias) event sample is
collected using coincidences between trigger signals from the $+z$ and $-z$ sides of either the scintillator counter counter(BSC) or the hadronic forward calorimeters(HF). 
The total integrated luminosity corresponding to our event sample amounts to 6.8 mb$^{-1}$.
\begin{figure}[htb]
  \begin{center}
    \resizebox{0.48\textwidth}{!}{\includegraphics{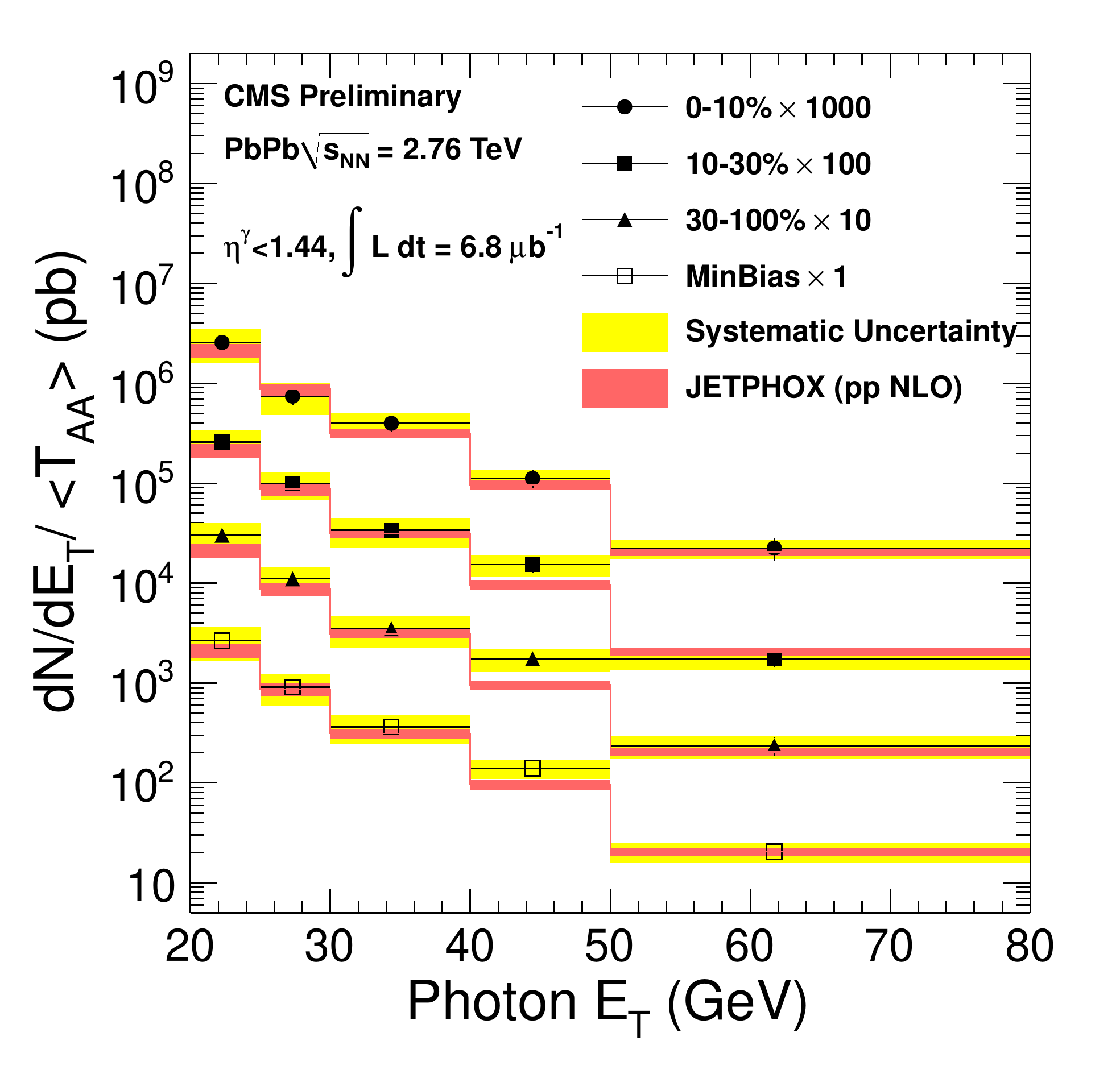}}
    \resizebox{0.48\textwidth}{!}{\includegraphics{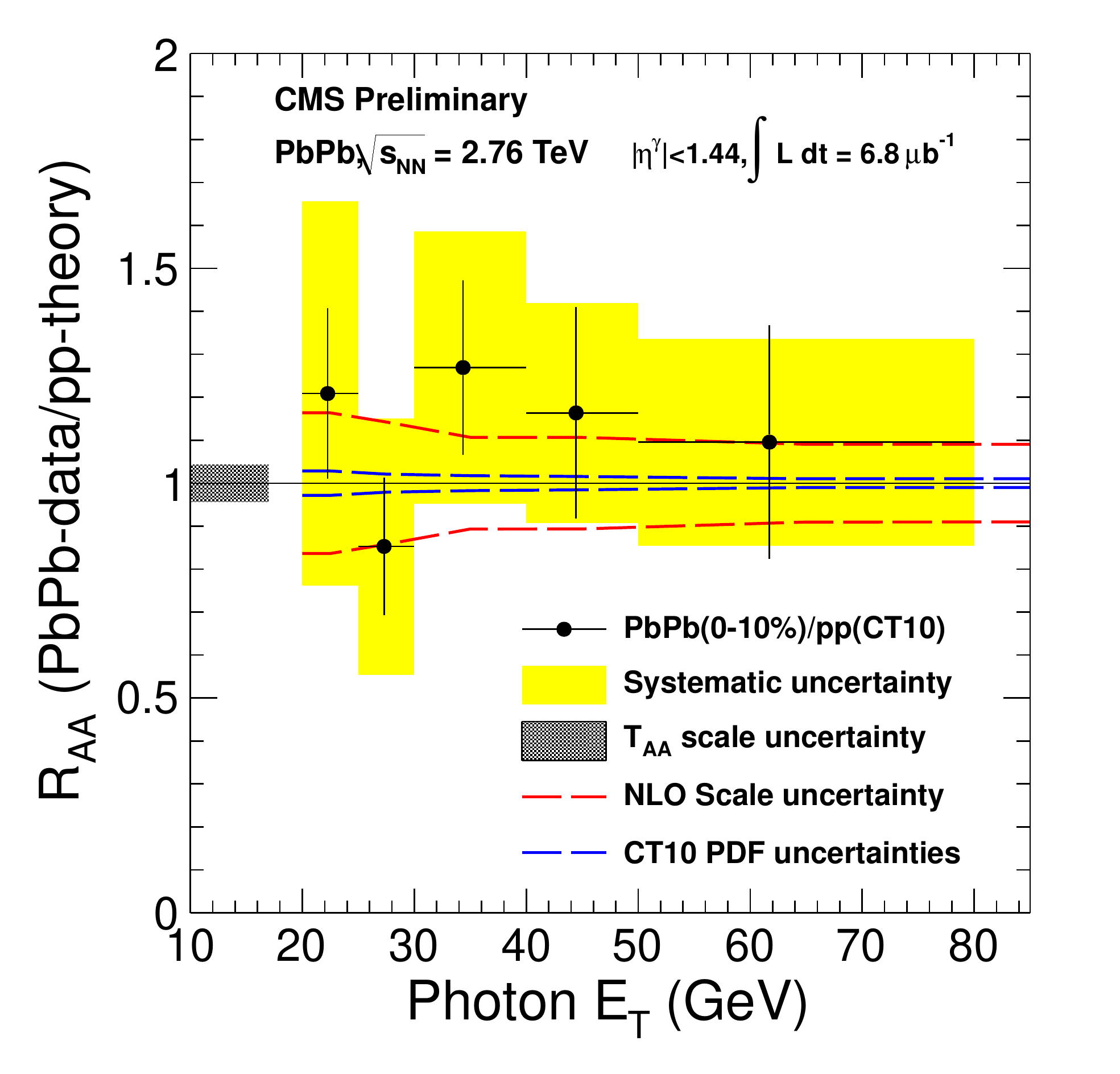}}
    \caption{\label{fig:dndet} (Left Panel)Normalized isolated photon yields $(dN^\gamma/d E_{T})/T_{AA}$ in each centrality intervals and MinBias events~\cite{HIN11002}.  (Right Panel)Nuclear modification factor $R_{AA}$ as a function of the photon $E_{T}$ measured in 0-10\% central PbPb collisions over the $T_{AA}$-scaled pp JETPHOX prediction at $\sqrt{s_{NN}} = 2.76$~TeV.  The dashed lines are uncertainties from CT10 PDF ~\cite{CT10} and NLO scale.
    }
  \end{center}
\end{figure}

In order to study the photon efficiency and electron rejection,
$\gamma$-jet, dijet and $W\rightarrow e\nu$ events are generated by PYTHIA generator (version 6.422, tune D6T)~\cite{Sjostrand:2006za}, modified
for the isospin of the colliding nuclei~\cite{Lokhtin:2005px}, and embedded in the MinBias data.

To determine whether a given photon is isolated in the generator level, we
define an isolation cone of $\Delta R = \sqrt{\Delta\phi^2+\Delta\eta^2}  < 0.4$ around its position in
pseudo-rapidity and azimuth. A photon is considered to be isolated if the sum of the particles
$p_T$ produced from the same hard scattering inside the isolation cone is smaller
than 5 GeV. 
And the centrality is determined with the MinBias sample using the total sum of energy signals from both positive and negative
HF (covering $2.9<|\eta|<5.2$).

The Island algorithm~\cite{Bayatian:2006zz} is used for the electromagnetic calorimeter (ECAL) energy clustering for photon reconstruction, and the energy is corrected taking into account the material in front of the ECAL and the electromagnetic shower containment.  An additional energy correction is applied to remove the background contribution from the underlying PbPb event.

The photon candidates are required to be 
in $|\eta^\gamma|<1.44$ and not to match with electron candidates
To reject candidates originated from jets, the energy ratio of hadronic calorimeter (HCAL) over ECAL inside a cone of $\Delta R = 0.15$ is required to be smaller than 0.2. 
The calorimeter-based isolation variables $\rm {Iso_{ECAL}}$ and $\rm{Iso_{HCAL}}$
are calculated by summing over the transverse energy measured inside the cone by the ECAL and HCAL, while the
track-based isolation variable $\rm {Iso_{Track}}$ is calculated by summing over the transverse momentum of
the tracks with $p_T>2$ GeV/$c$.  
In order to remove the contribution of hadronic activity from the uncorrelated underlying PbPb event background, the average value of the given cone variable per unit area in the $\eta-\phi$ phase space($\rm  \langle Iso \rangle$) is estimated in a rectangular area, $2\Delta R$ wide centred at $\eta^\gamma$ in the $\eta$-direction and  2$\pi$ in the $\phi$-direction, excluding the isolation cone. 
The sum of the isolation variables $(\rm {SumIso = Iso'_{ECAL} + Iso'_{HCAL} + Iso'_{Track}})$ is required to be smaller than 5 GeV.

Remaining backgrounds are estimated using the template method.
The shape variable $\sigma_{i\eta i \eta}$ is used to characterize the shower shape, which is defined as:
\begin{eqnarray}
  \label{sieieFormula}
  \sigma_{i\eta i\eta}^2 = \frac{\sum_i w_i(\eta_i-\bar{\eta})^2}{\sum_i w_i}, w_i = {\rm max}(0, 4.7 + \ln \frac{E_i}{E}),
\end{eqnarray}
where $E_i$ and $\eta_i$ are the energy and position of the $i^{th}$ crystal in a group of $5\times 5$ centred in the highest energy one. 
The photon candidates tend to have smaller $\sigma_{i\eta i\eta}$ while hadrons and $\pi^0$s tend to have larger $\sigma_{i\eta i \eta}$.

The probability distribution function of this shower shape variable is called a template.
Once we have the templates of signal and background, we can extract the number of signals in data by fitting the linear combination of two templates into the shower shape distribution of data. The signal template is determined from $\gamma$-jet PYTHIA+MB samples. The template for background, i.e. non-photon events, is obtained from the data using a side-band shifted in SumIso ($6<{\rm SumIso}<11$ GeV).

The systematic uncertainty of the photon yield $dN/dE_{T}$ is dominated by the uncertainty of the background
template.
Non-prompt photons from PYTHIA + MB samples are used to monitor the difference between the templates in signal and side-band regions.
The total systematic uncertainties are found to be 21-37\%.

\begin{figure}[htb]
  \begin{center}
    \resizebox{0.62\textwidth}{!}{\includegraphics{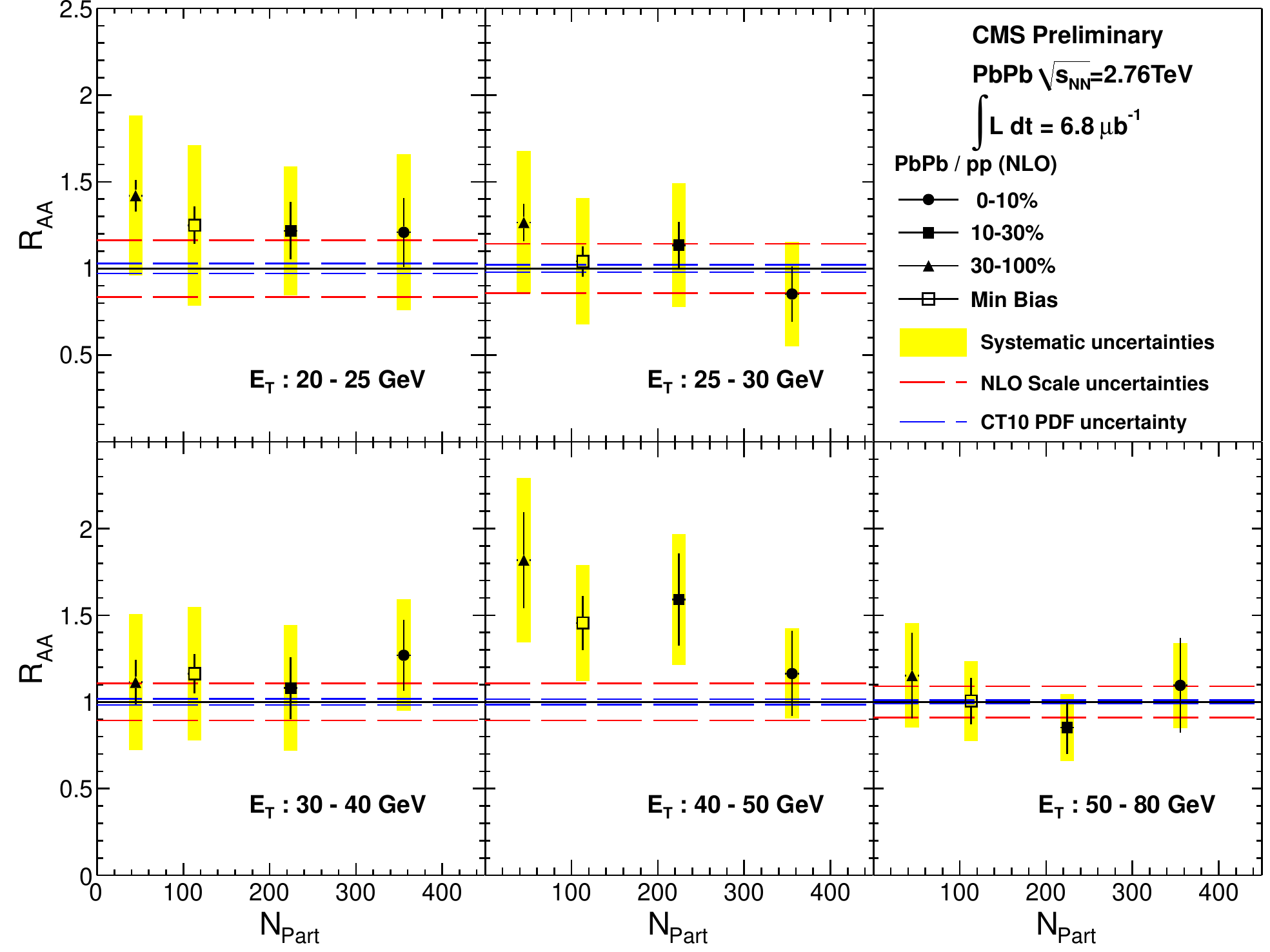}}
    \caption{\label{fig:RAAvsNpart} Nuclear modification factor $R_{AA} $ as a function of $N_{\rm part}$ for the five different photon transverse energy intervals~\cite{HIN11002}.
    }
  \end{center}
\end{figure}

\section{Results}
The data are compared with NLO pQCD predictions from JETPHOX 1.2.2 ~\cite{JETPHOX}, using  the CT10 PDF ~\cite{CT10} and the BFG set II of fragmentation function ($ff$) ~\cite{BFGII}.

Scaling factor {\it nuclear overlap function} ($T_{AA}$) was obtained from Glauber model calculation~\cite{Miller:2007ri} to provide proper normalization. 
The left panel of Figure ~\ref{fig:dndet} shows the normalized yields $(dN^\gamma/d E_{T})/T_{AA}$ and pp prediction.
The {\it nuclear modification factor} $R_{AA}= dN^\gamma/(T_{AA}\times \sigma^\gamma_{pp})$, is computed from the PbPb measured yield $dN^\gamma$, the nuclear overlap function $T_{AA}$ and the inclusive isolated photon cross-section $\sigma^\gamma_{pp}$ given by the JETPHOX calculation.
The right panel of Figure ~\ref{fig:dndet} shows the $R_{AA}$ as a function of the photon $E_T$ in the 0-10\% central collisions. The results are found to be compatible with unity within the quoted uncertainties. 
In addition, Figure ~\ref{fig:RAAvsNpart} shows the $R_{AA}$ as a function of $N_{\rm part}$. No significant centrality dependence is observed in data. 

\section{Summary}
The isolated photon yield with $|\eta|<1.44$ in PbPb collisions at $\sqrt{s_{NN}} = 2.76$ TeV
has been measured as a function of the transverse momentum. No modification is observed with respect to the
next-to-leading order calculation from JETPHOX scaled by the number of incoherent nucleon-nucleon collisions based on a Glauber model calculation. This also establishes the base of the future analysis based on photons as unmodified hard probes in the studies of the produced medium in the PbPb collisions at the LHC.

\bibliography{auto_generated}   

\providecommand{\href}[2]{#2}\begingroup\raggedright\begin{thebibliography}{10}%
\makeatletter
\providecommand{\hrefCMSnoop }[0]{\@secondoftwo}%
\makeatother

\bibitem{Arleo:2011gc}
\hrefCMSnoop {} {{Arleo, Francois and Eskola, Kari J. and Paukkunen, Hannu and
  Salgado, Carlos A.}, ``{Inclusive prompt photon production in nuclear
  collisions at RHIC and LHC}'',} \textit{ JHEP} \textbf{ 04} (2011) 055,
  \href{http://www.arXiv.org/abs/1103.1471}{\texttt{ arXiv:1103.1471}}.
\href{http://dx.doi.org/10.1007/JHEP04(2011)055}{\texttt{
  doi:10.1007/JHEP04(2011)055}}.

\bibitem{JINST}
\hrefCMSnoop {} {{ CMS} Collaboration, ``{The CMS experiment at the CERN
  LHC}'',} \textit{ JINST} \textbf{ 3} (2008) S08004.
\href{http://dx.doi.org/10.1088/1748-0221/3/08/S08004}{\texttt{
  doi:10.1088/1748-0221/3/08/S08004}}.

\bibitem{HIN11002}
\href {http://cdsweb.cern.ch/record/1352779?ln=en} {{ CMS} Collaboration,
  ``{Isolated photon production in $\sqrt{s_{_{NN}}} = 2.76$ TeV PbPb
  collisions as a function of transverse energy and reaction centrality}'',}
  \textit{ CMS Physics Analysis Summary} \textbf{ CMS-PAS-HIN-11-002} (2011).

\bibitem{CT10}
\hrefCMSnoop {} {{H.-L. Lai, M. Guzzi, J. Huston et al.}, ``{New parton
  distributions for collider physics}'',} \textit{ Phys. Rev.} \textbf{ D82}
  (2010) 074024, \href{http://www.arXiv.org/abs/1007.2241}{\texttt{
  arXiv:1007.2241}}.
\href{http://dx.doi.org/10.1103/PhysRevD.82.074024}{\texttt{
  doi:10.1103/PhysRevD.82.074024}}.

\bibitem{Sjostrand:2006za}
\hrefCMSnoop {} {T.~Sjostrand, S.~Mrenna, and P.~Z. Skands, ``{PYTHIA 6.4
  Physics and Manual}'',} \textit{ JHEP} \textbf{ 05} (2006) 026,
  \href{http://www.arXiv.org/abs/hep-ph/0603175}{\texttt{
  arXiv:hep-ph/0603175}}.
\href{http://dx.doi.org/10.1088/1126-6708/2006/05/026}{\texttt{
  doi:10.1088/1126-6708/2006/05/026}}.

\bibitem{Lokhtin:2005px}
\hrefCMSnoop {} {I.~P. Lokhtin and A.~M. Snigirev, ``{A model of jet quenching
  in ultrarelativistic heavy ion collisions and high-p(T) hadron spectra at
  RHIC}'',} \textit{ Eur. Phys. J.} \textbf{ C45} (2006) 211,
  \href{http://www.arXiv.org/abs/hep-ph/0506189}{\texttt{
  arXiv:hep-ph/0506189}}.
\href{http://dx.doi.org/10.1140/epjc/s2005-02426-3}{\texttt{
  doi:10.1140/epjc/s2005-02426-3}}.

\bibitem{Bayatian:2006zz}
\hrefCMSnoop {} {{ CMS} Collaboration, ``{CMS physics: Technical design
  report}'',}. CERN-LHCC-2006-001.

\bibitem{JETPHOX}
\hrefCMSnoop {} {{S. Catani, M. Fontannaz, J.P.Guillet et al.}, ``{Cross
  section of isolated prompt photons in hadron-hadron collisions}'',} \textit{
  JHEP} \textbf{ 05} (2002) 028,
  \href{http://www.arXiv.org/abs/hep-ph/0204023}{\texttt{
  arXiv:hep-ph/0204023}}.
  \href{http://dx.doi.org/10.1088/1126-6708/2002/05/028}{\texttt{
  doi:10.1088/1126-6708/2002/05/028}}.

\bibitem{BFGII}
\hrefCMSnoop {} {{L. Bourhis, M. Fontannza, and J. P. Guillet}, ``{Quark and
  gluon fragmentation functions into photons}'',} \textit{ Eur. Phys. J.}
  \textbf{ C2} (1998) 529, \href{http://www.arXiv.org/abs/9704447}{\texttt{
  arXiv:9704447}}. \href{http://dx.doi.org/10.1007/s100520050158}{\texttt{
  doi:10.1007/s100520050158}}.

\bibitem{Miller:2007ri}
\hrefCMSnoop {} {{Miller, Michael L. and Reygers, Klaus and Sanders, Stephen J.
  and Steinberg, Peter}, ``{Glauber modeling in high energy nuclear
  collisions}'',} \textit{ Ann. Rev. Nucl. Part. Sci.} \textbf{ 57} (2007) 205,
  \href{http://www.arXiv.org/abs/nucl-ex/0701025}{\texttt{
  arXiv:nucl-ex/0701025}}.
\href{http://dx.doi.org/10.1146/annurev.nucl.57.090506.123020}{\texttt{
  doi:10.1146/annurev.nucl.57.090506.123020}}.

\end{thebibliography}\endgroup

\end{document}